\documentclass[%
 reprint,
]{revtex4-2}
\usepackage[colorlinks,linkcolor=blue,urlcolor=blue,citecolor=blue]{hyperref}
\usepackage{graphicx}
\usepackage{dcolumn}
\usepackage{amsmath}
\usepackage{bm}
\usepackage{subfigure}
\usepackage{soul}
\usepackage{xcolor} 
\begin{document}

\preprint{APS/123-QED}

\title{Multi-color x-ray free-electron laser generation using optical klystron for multi-frame diffraction imaging}

\author{Xiaodan Liu\textsuperscript{1,}\textsuperscript{2}}

\author{Hanxiang Yang\textsuperscript{2}}
\author{Bingyang Yan\textsuperscript{3,}\textsuperscript{4}}
\author{Yue Wang\textsuperscript{5}}
\author{Nanshun Huang\textsuperscript{2}}
\author{Liqi Han\textsuperscript{6}}
\author{Jie Cai\textsuperscript{6}}
\author{Han Wen\textsuperscript{1}}
\author{Jinqing Yu\textsuperscript{1,}}%
\thanks{jinqing.yu@hnu.edu.cn}
\author{Haixiao Deng\textsuperscript{2,}}%
\thanks{denghx@sari.ac.cn}
\author{Xueqing Yan\textsuperscript{6,}}%
\thanks{x.yan@pku.edu.cn}
\affiliation{\textsuperscript{1}Hunan Provincial Key Laboratory of High-Energy Scale Physics and Applications, School of Physics and Electronics, Hunan University, Changsha 410082, China\\          
    \textsuperscript{2}Shanghai Advanced Research Institute, Chinese Academy of Sciences, Shanghai 201210, China\\
    \textsuperscript{3}Shanghai Institute of Applied Physics, Chinese Academy of Sciences, Shanghai 201800, China\\
	\textsuperscript{4}University of Chinese Academy of Sciences, Beijing 101408, China\\
	\textsuperscript{5}Zhangjiang Laboratory, Shanghai 201210, China\\
	\textsuperscript{6}State Key Laboratory of Nuclear Physics and Technology, and Key Laboratory of HEDP of the Ministry of Education, CLAPA, Peking University, Beijing 100871, China
}%

\date{\today}
\begin{abstract}
X-ray free-electron lasers (XFELs) of high brightness have opened new opportunities for exploring ultrafast dynamical processes in matter, enabling imaging and movies of single molecules and particles at atomic resolution. In this paper, we present a straightforward method for multi-frame diffraction imaging, using the whole electron beam to generate four-color XFEL pulses with adjustable wavelength separation and time delay. The optical klystron scheme is introduced to enhance FEL intensity and reduce the total length of undulators. The time delay is tuned via a magnetic chicane between the undulators with various colors. Using parameters of SHINE, start-to-end simulations demonstrate the effectiveness and tunability of our method, achieving representative results such as time delays of hundreds of femtoseconds and four-color XFEL pulses spanning 1.8 to 2.7 nm with 0.3 nm intervals. The proposed scheme enables the recording of multi-frame diffraction images in a single exposure, providing a new perspective for ultrafast molecular and atomic dynamics studies.
\end{abstract}

\maketitle

\section{\label{sec1}Introduction}
In recent years, X-ray free-electron lasers (XFELs) \cite{2007huangReviewXrayFreeelectron,2016pellegriniPhysicsXrayFreeelectron,2021Features} have emerged as transformative tools for probing ultrafast dynamics in matter, owing to their femtosecond-to-attosecond pulse durations, high brightness, and full transverse coherence. In time-resolved studies, pump-probe configurations that employ either a conventional optical laser or an FEL-generated X-ray pump combined with an X-ray FEL probe are indispensable for capturing ultrafast dynamics across physical phenomena \cite{2012materipfauUltrafastOpticalDemagnetization,2013allariaTwocolourPumpProbe,2016Observation,2016plasmagorkhoverFemtosecondNanometreVisualization,2016ferrariWidelyTunableTwocolour,2018densepontiusProbingNonequilibriumTransienta,2021chenobservation,2023plasmahoeingTimeresolvedSingleparticleXraya}, chemical reactions \cite{2017chemattarFemtosecondXraySpectroscopya,2018chemincigrucciImpulsiveUVpumpXraya,2019berrahFemtosecondresolvedObservationFragmentation}, and biological structures \cite{2011biocalemanSimulationsRadiationDamagea,2017biowolfProbingUltrafastPpa} on the picosecond to femtosecond time scale. However, these methods rely on shot-to-shot repetitive measurements with scanned pump-probe delays for image construction, which limits their application to non-repetitive, irreversible, or poorly reproducible events \cite{2014nakagawaSequentiallyTimedAlloptical}. In such cases, it is highly desirable to record multiple transient states originating from the same excitation event in a single shot, analogous to high-speed photography or flash imaging \cite{2023saikiSingleshotOpticalImaging}. Therefore, single-shot multi-color XFEL pulses with well-defined and controllable relative time delays provide an effective approach for enabling multi-frame diffraction imaging of ultrafast dynamics.

Various multi-color XFEL schemes have been proposed by manipulating electron beam properties, including twin bunches \cite{2015marinelliHighintensityDoublepulseXray}, two-bucket bunches \cite{2022deckerTunableXrayFree}, and the use of different energy parts within one bunch \cite{2021s.bettoniExperimentalDemonstrationTwocolor}, double-slot foil \cite{2019saahernandezGenerationTwocolorXray}, sextupole \cite{2020dijkstalDemonstrationTwocolorXray}, nonlinear electron compression \cite{2020malyzhenkovSingleTwocolorAttosecond} and laser emittance spoiler \cite{2021vicarioTwocolorXrayFreeelectron}. Meanwhile, artificially changing the undulator parameter can straightforwardly produce tunable multi-color XFEL pulses \cite{2013lutmanExperimentalDemonstrationFemtosecond,2013haraTwocolourHardXray,2023choGenerationTimesynchronizedTwocolor}. However, when the whole electron bunch is used to generate multiple colors sequentially, the intensity of downstream pulses is limited by the energy spread induced by upstream lasing. The fresh-slice technique circumvents this limitation by selectively utilizing unmodulated portions of the beam for independent amplification in separate undulators, thereby enabling high-power multi-color operation \cite{2016lutmanFreshsliceMulticolourXray,2016reicheTwocolorOperationFreeelectron,2018guetgDispersionBasedFreshSliceScheme,2018chaoControlLasingSlice,2022pratWidelyTunableTwocolor,2024Millijoule}.

Inspired by a single-shot multi-frame imaging concept previously demonstrated at optical wavelengths \cite{2023saikiSingleshotOpticalImaging}, we propose an extension to XFELs. In contrast to fresh-slice schemes, the proposed scheme generates four-color XFEL pulses from the whole electron bunch, enabling single-shot multi-frame X-ray diffraction imaging with adjustable wavelength separation and time delay. To mitigate the constraints imposed by XFEL-induced energy spread and undulator length, we employ an optical klystron (OK) scheme \cite{1992bonifacioTheoryHighgainOptical,2001Dispersively,2003saldinFreeElectronLaser,2015pencoExperimentalDemonstrationEnhanced,2017pencoOpticalKlystronEnhancement,2021Development}, which enhances gain and reduces saturation length, facilitating high-power multi-color pulse generation with straightforward implementation in existing facilities. The proposed scheme can generate four discrete-wavelength XFEL pulses with tunable wavelength separation and time delay, and the pulse energy of each color can be adjusted using the $R_{56}$ of the dispersive chicane. Moreover, the feasibility of the proposed scheme is confirmed by the reflection grating diffraction results. In operation, a single pump pulse initiates a reaction, followed by three or four XFEL probe pulses with femtosecond- to picosecond-scale delays. These tailored XFEL pulses sequentially probe the same reaction. The resulting diffraction signals are spatially separated by a grating and simultaneously recorded on an X-ray CCD detector. Since all pulses originate from the same electron bunch, they are intrinsically synchronized, which reduces relative timing uncertainties and thereby enables single-shot, multi-frame time-resolved flash imaging of ultrafast dynamics.

This paper is organized as follows. Section~\ref{sec2} elaborates on the principles of the proposed scheme. Numerical simulations of multi-color XFEL generation and transport based on the main parameters of Shanghai High Repetition Rate XFEL and Extreme Light Facility (SHINE) are presented in Section~\ref{sec3} and ~\ref {sec4}. Conclusion and discussion are described in Section~\ref{sec5}.

\section{\label{sec2}Principles}
The proposed scheme can be divided into two phases, as shown in Fig.~\ref{fig:figure-scheme}. First, the generation of multi-color XFEL pulses employs the electron beam passing through these undulator modules ($A_{N}-D_{N}$, each with distinct undulator parameters $K$), and three time-delay chicanes as shown in Fig.~\ref{fig:figure-scheme}(a) and  Fig.~\ref{fig:figure-scheme}(b). Second, the high-time-resolution dynamic diffraction imaging system features the four-color X-ray pulses interacting with a sample, followed by a high line-density reflective grating, and an X-ray CCD detector, as depicted in Fig.~\ref{fig:figure-scheme}(c).

\begin{figure*}[bt]
\includegraphics[width=0.52\textwidth]{fig1.pdf}
\caption{\label{fig:figure-scheme}Proposed multi-frame X-ray diffraction imaging scheme. (a) Schematic layout of the multi-color XFEL pulses generation. The setup consists of an electron beam, four undulator modules denoted as $A_{N}$, $B_{N}$, $C_{N}$ and $D_{N}$, three time-delayed chicanes, and an electron beam collector, where N denotes the number of sub-undulators in each module. The modules generate XFEL pulses at wavelengths $\lambda_1-\lambda_4$, color-coded by wavelength. The dashed box indicates an undulator module detailed in panel (b), which comprises N sub-undulators $U_{X,i} \ (X=A,B,C,D;\ i=1,\dots,N)$,  interleaved with N-1 small dispersive chicanes. (c) Schematic layout of the high-time-resolution dynamic diffraction imaging system. This includes a sample, a high line-density reflective grating, and an X-ray CCD detector. The four X-ray pulses sequentially illuminate the sample to capture temporal information of ultrafast events, which are dispersed spatially via the grating and recorded as four diffraction images by the CCD detector.}

\end{figure*}

As the electron beam propagates through sub-undulator $U_{A,1}$, FEL radiation is generated at the resonance wavelength, given by \cite{2007huangReviewXrayFreeelectron}
\begin{equation}
\label{resonance}
\lambda_r = \frac{\lambda_u}{2\gamma^2} \left( 1 + \frac{K^2}{2} \right)
\end{equation}
where $\lambda_u$ denotes the undulator period, $K$ represents the undulator parameter, and $\gamma$ is the Lorentz factor. In the amplification process of the OK-enhanced self-amplified spontaneous emission (SASE) scheme \cite{2017pencoOpticalKlystronEnhancement}, the energy modulation is converted to density modulation via a small dispersive chicane. Moreover, this configuration speeds the electron beam's microbunching, thereby reducing the FEL gain length. Notably, the OK efficiency is substantially enhanced when the electron beam’s rms relative energy spread $\sigma_\delta$ is much smaller than the FEL Pierce parameter $\rho$ \cite{2007huangReviewXrayFreeelectron}. The one-dimensional theory describing the maximum theoretical power gain factor in the OK-SASE occurs when \cite{2006Optical}:
\begin{equation}
\label{condition}
\frac{2\pi}{\lambda_r} R_ {56} \sigma _ { \delta } \sim 1, 
\end{equation}
where $R_ {56}$ denotes the momentum compaction of the small dispersive chicane. However, if the first sub-undulator $U_{A,1}$ is sufficiently long, it can generate an energy modulation with an amplitude that is comparable to or exceeds the intrinsic energy spread. In this scenario, Equation~\eqref{condition} no longer corresponds to the maximum gain, as the induced energy modulation becomes dominant over the intrinsic energy spread.
By considering the high-gain harmonic generation \cite{1991yuGenerationIntenseUv}, we can analyze the scenario that may arise in the aforementioned situation \cite{2017pencoOpticalKlystronEnhancement}. The nth harmonic bunching factor can be derived as:
\begin{equation}
\label{bunching}
b _ { n } =  e x p ( - \frac { 1 } { 2 } n ^ { 2 } \sigma _ { \delta } ^ { 2 } k _ { r } ^ { 2 } R _ { 5 6 } ^ { 2 } )| J _ { n } ( n \frac { \Delta E } { E } k _ { r } R _ { 5 6 } ) | ,
\end{equation}
where $J_{n}$ is the $n$th order Bessel function, $\Delta E$ denotes the energy modulation amplitude induced by the first sub-undulator $U_{A,1}$ and $E$ represents the mean energy of the electron bunch.  From Equation~\eqref{bunching}, we can analyze the OK behavior of $R_{56}$ to the effects of different energy modulation amplitudes, specifically $\Delta E $ = 250 keV and $\Delta E$ = 2000 keV,  for an electron bunch with a mean energy of $E$ = 4.5 GeV and an intrinsic energy spread of 650 keV ($\sigma _ { \delta } =1.44\times10 ^{-4} $). Furthermore, the undulator module $A_N$ is tuned to resonate at a wavelength of 1.8 nm, and only the fundamental harmonic ($n = 1$) is considered.
Figure~\ref{fig:figure_bunching}(a) illustrates the scenario where the relative energy modulation amplitude is smaller than the rms relative intrinsic energy spread, i.e., $\frac { \Delta E } { E }< \sigma _ { \delta }$,  the contribution of the exponential decay to the bunching factor becomes more significant than that of the Bessel function. In this case, the value of $R_{56}^1$ corresponding to the maximum bunching is found to be $ 1.94 \ \mu m$. Notably, this value matches optimal $R_{56}^1$ calculated from Eq.~\eqref{condition}, which also yields $ 1.98 \ \mu m$. This agreement confirms that the bunching factor reaches its maximum when the condition described by Eq.~\eqref{condition} is satisfied. In contrast, as shown in Fig.~\ref{fig:figure_bunching}(b), when the Bessel function dominates, the optimal value of $R_{56}^1$ shifts to $1\ \mu m$, where the maximum bunching factor is achieved under the condition $\frac { \Delta E } { E } k _ { r } R _ { 5 6 } \approx 1.8412$. This highlights the transition between the dominance of the exponential decay and the Bessel function in determining the bunching behavior, with each regime yielding distinct optimal values of $R_{56}^1$ corresponding to their respective conditions for maximizing the bunching factor.

\begin{figure*}[tb]
\includegraphics[width=0.48\textwidth]{fig2.pdf}
\caption{\label{fig:figure_bunching}The square of the bunching factor (shown in black) is plotted as a function of $R_{56}^1$ for the first small dispersive chicane at the fundamental harmonic. The bunching factor is obtained through the combined contributions of an exponential decay (in red) and a Bessel function (in blue). Two cases are considered, corresponding to energy modulation amplitudes of 250 keV (a) and 2000 keV (b).}
\end{figure*}

\begin{figure}[tb]
\includegraphics[width=0.26\textwidth]{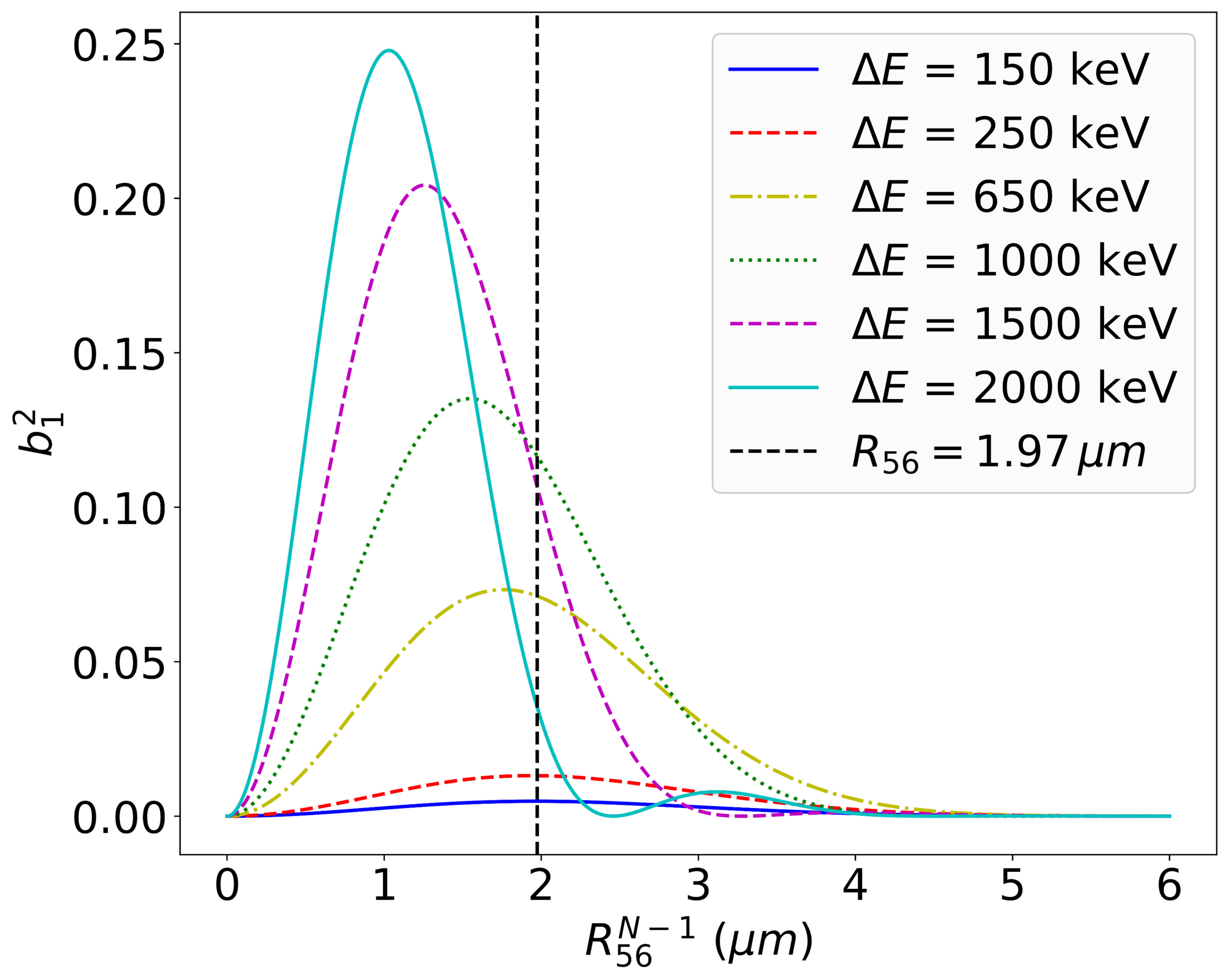}
\caption{\label{fig:figure_dif_ener}The different energy modulation amplitudes induced by the sub-undulator $A_{N-1}$ affect the square of the bunching factor, which is related to $R_{56}^{N-1}$. The vertical plot corresponds to the optimal $R_{56}$ value that satisfies the Eq.~\eqref{condition}.}
\end{figure}

When the bunched electron beam passes through the sub-undulator $U_{A,N-1}$, it undergoes $N-2$ stages of OK-enhanced SASE FEL amplification, specifically referred to as the seeded FEL amplification. During this process, the energy modulation amplitude induced by the sub-undulator $U_{A,N-1}$ before the last dispersive chicane $R_{56}^{N-1}$ becomes larger than the intrinsic energy spread. As the energy modulation amplitude increases, the optimal value of $R_{56}^{N-1}$ decreases due to the dominance of the Bessel function governing the bunching process. Specifically, a larger energy modulation amplitude requires a smaller value $R_{56}^{N-1}$ to maximize the bunching factor, as shown in Fig.~\ref{fig:figure_dif_ener}. This optimization ultimately enhances FEL radiation in the subsequent sub-undulator $U_{A,N}$. 
Additionally, the OK enhancement is also highly sensitive to the initial slice energy spread \cite{1992bonifacioTheoryHighgainOptical,2006Optical,2021Development}. As shown in Fig.~\ref{fig:re-beam-espread}, the pulse energy decreases rapidly as the initial slice energy spread increases, indicating clear performance degradation. For each point, the corresponding $R_{56}$ values were re-optimized using the differential evolution algorithm \cite{2025BaejDetPy} to maximize the pulse energy at $\lambda_1$. 
\begin{figure}[tb]
\includegraphics[width=0.26\textwidth]{re-beam-espread.pdf}
\caption{\label{fig:re-beam-espread}Dependence of the pulse energy at 1.8 nm on the initial slice energy spread. The beam energy and peak current are fixed at 4.5 GeV and 800 A, respectively.}
\end{figure}
The time delay of delay chicane 1 is twice its $R_{56}$, which is sufficient to smear out the microbunching generated by the previous undulator module $A_N$. For instance, when $R_{56} = 600 \ \mu m$, the bunching factor is nearly zero, as given by Eq.~\eqref{bunching}. Similarly, the multi-stage OK configuration involves $N-1$ small dispersive chicanes for each wavelength, including $\lambda_2$, $\lambda_3$, and $\lambda_4$.

Therefore, four-color XFEL pulses with distinct wavelengths ($\lambda_1 < \lambda_2 < \lambda_3 < \lambda_4$) and temporal intervals between adjacent pulses ranging from picoseconds to femtoseconds provide a novel experimental methodology for ultrafast diffraction experiments utilizing multiple monochromatic short pulses in Fig.~\ref{fig:figure-scheme}(c). When a set of four-color pulses is generated within a single exposure and sequentially interacts with an object, and the pulses are diffracted by a grating with high line-density. At the grating's image plane, diffracted light from pulses of different colors ultimately falls at distinct positions. Consequently, information carried by different pulses within the set is recorded at separate image plane locations. Furthermore, the temporal delay between pulses enables a four-color pulse train to comprehensively capture material information over a specific time interval. Upon passing through a high line-density reflective grating \cite{2021Soft}, the X-rays can be spatially separated according to the grating equation:
\begin{equation}
\label{grating}
d \left( \sin\theta + \sin\theta' \right) = m\lambda,
\end{equation}
where $d$ denotes the grating period, $\theta$ is the incident angle, $\theta'$ is the diffracted angle, $m$ is the diffraction order and $\lambda$ is the FEL wavelength. To analyze the angular dispersion of the grating, we keep the incident angle $\theta$ constant and differentiate both sides of the grating Eq.~\eqref{grating} with respect to $\lambda$. This yields the following:
\begin{equation}
\frac{d\theta'}{d\lambda} = \frac{m}{d \cos\theta'},
\end{equation}
the resolving power of the grating is determined by the relationship $R = \frac{\lambda}{\Delta\lambda_{min}} = m N$, where $\Delta\lambda_{min}$ denotes the minimum resolvable wavelength difference and $N$ is the total number of illuminated grooves on the grating, given by $N = \frac{L}{d}$, with $L$ being the projected length of the beam on the grating surface. For the four-color XFEL pulses, the full width at half maximum (FWHM) bandwidth of each wavelength was greater than twice its specific minimum resolvable wavelength difference, and the wavelength difference between adjacent pulses is sufficiently large such that it exceeds the minimum resolvable wavelength differences. As a result, the X-rays are clearly separated by the grating, spatially dispersed in ascending order of wavelength ($\lambda_1<\lambda_2<\lambda_3<\lambda_4$), and symmetrically arranged around the zeroth-order diffraction peak. Each X-ray pulse generates a unique diffraction angle via the grating, resulting in angles $\theta_1'$, $ \theta_2'$, $\theta_3'$, and $\theta_4'$, allowing for clear and distinct separation of the four X-ray pulses.

The X-ray CCD detector records these diffracted images: the first image, formed by the initial X-ray pulse, corresponds to diffraction angle $\theta_1'$ and time $t_1$. Subsequently, within the picosecond-to-femtosecond timeframe, the second, third, and fourth pulses produce images at angles $\theta_2'$, $\theta_3'$, and $\theta_4'$ (times $t_2$, $t_3$, and $t_4$, respectively). Therefore, X-rays of different wavelengths carry distinct temporal information of ultrafast events at the corresponding time points. The four-color X-ray pulses are then spatially separated via wavelength-to-spatial dispersion using a grating, and the resulting diffraction patterns are recorded by an X-ray CCD detector. This yields a series of multi-frame diffraction images with inherent temporal correlation, enabling the reconstruction of ultrafast dynamical processes.   

In the proposed scheme, adjusting the time delay of the large chicane controls the inter-pulse time separation between adjacent X-ray pulses to set the frame interval. Tuning the $K$ parameter of sub-undulators regulates both the wavelength separation and specific X-ray wavelengths. The energy of each X-ray pulse can be effectively controlled by jointly tuning the number of sub-undulators and the optimized $R_{56}$ values of dispersive chicanes. Theoretically, this scheme enables the generation of multi-color X-ray pulses for multi-frame diffraction imaging.  However, limitations in adjustable frame numbers
exist due to electron beam energy spread growth and the dynamic range constraints of the X-ray CCD, which are further influenced by its physical size.

\section{\label{sec3}Start-to-end XFEL simulation}
In this section, we explore the generation of multi-color XFEL pulses at the soft X-ray beamline of SHINE \cite{2023SHINE}. The SHINE employs an 8 GeV CW superconducting RF linac \cite{2019yanMultibeamenergyOperationContinuouswave,2025chenultra} to deliver X-ray photons from 0.4-25 keV at a 1 MHz repetition rate, with two FEL beamlines: hard X-ray (FEL-I) and soft X-ray (FEL-II). Besides 8 GeV full beam energy operation, FEL-II beamline \cite{2023SHINE} can operate with a 3-4.5 GeV electron beam and support various modes, including SASE, self-seeding, and external seeded FEL schemes at MHz repetition rate \cite{2021yanself,2025qifirst}. Its undulator system, featuring a 55 mm period length, comprises 32 undulator modules, each 4 m long. Potentially, small dispersive chicanes can be integrated within the 1 m interspaces between these modules, thereby forming reconfigured functional modules.

The start-to-end (S2E) simulation was performed using ASTRA \cite{2004Astra} for the electron beam dynamics in the photoinjector, ELEGANT \cite{2001elegant} for the linac transport including coherent synchrotron radiation effects, and Genesis 1.3 \cite{S1999GENESIS} for the three-dimensional FEL simulations. The electron beam, whose current profile and longitudinal phase space properties are shown in Fig.~\ref{fig:twiss}, is characterized by the parameters listed in Table~\ref{table1}. In the multi-color FEL simulations, the particle distribution is transferred between modules, and wavelength switching is handled by reslicing the beam according to the new resonant wavelength.

\begin{figure}[bt]
\includegraphics[width=0.48\textwidth]{twiss.pdf}
\caption{\label{fig:twiss}The current, energy, energy spread, and emittance of the electron beam in numerical S2E simulations. The bunch head is on the right.}
\end{figure}

\begin{table}
\caption{\label{table1}Main electron beam and undulator parameters of FEL-II.}
\begin{ruledtabular}
\begin{tabular}{lcc}
Parameters	&Value	&Unit\\ \hline
{\textbf{\textit{Electron beam}}}	&\\Energy &4.5		&GeV\\
			Slice energy spread 	&0.01     	&\%\\
			Normalized emittance  	&0.15	        &mm$\cdot$mrad\\
			Bunch charge			&100	    &pC\\
			Peak current &800	    &A\\
{\textbf{\textit{Undulator line}}}	&\\Period length& 55  &mm\\
Undulator length & 4  &m\\
Dispersion chicane $R_{56}$ & 0.2-2  &$\mu$m\\
Delay chicane $R_{56}$ &  $\leq 600$   &$\mu$m\\
\end{tabular}
\end{ruledtabular}
\end{table}
Under fully optimized $R_{56}$ values for all dispersive chicanes, the undulator modules $A_{5}$, $B_{6}$, $C_{6}$, $D_{6}$ (with five sub-undulators in $A_{5}$ and six sub-undulators in each of $B_{6}$, $C_{6}$ and $D_{6}$) are operated with undulator parameters $K_1$ = 2.00, $K_2$ = 2.21, $K_3$ = 2.39 and $K_4$ = 2.56, as illustrated in Fig.~\ref{fig:figure_op}(c). The electron beam propagating through these undulator modules generates radiation at the resonant wavelengths of 1.8, 2.1, 2.4, and 2.7 nm,  with adjacent wavelengths separated by 0.3 nm, shown in Fig.~\ref{fig:figure_op}(a). These wavelengths correspond to photon energies of 688.80, 590.40, 516.60, and 459.20 eV, respectively. 
The four-color X-ray pulses achieve peak power levels on the order of hundreds of megawatts, as illustrated in Fig.~\ref{fig:figure_op}(b). Independent generation of each X-ray pulse is enabled by delay chicanes, which provide a temporal separation of 1 ps between neighboring pulses, corresponding to the maximum delay used to clearly separate the four pulses. The minimum useful delay is constrained by the XFEL pulse length, below which neighboring pulses would overlap and become difficult to distinguish. This inter-pulse delay defines the accessible temporal sampling points between successive pulses, but it does not directly determine the temporal resolution. Instead, the temporal resolution is mainly limited by the XFEL pulse length and is on the order of 100 fs for the present case, consistent with the values listed in Table~\ref{table-ok/sase}. 
To realize this multi-color output within a single pass, the same electron bunch is reused across multiple undulator modules. However, the FEL radiation in the upstream sections induces significant energy spread in the high-current core of the bunch, thereby degrading its FEL gain for downstream wavelengths. Consequently, lasing at subsequent colors is sustained primarily by the low-energy-spread electrons in the bunch head and tail. This longitudinal partitioning of lasing regions leads to temporal pulse splitting in the output. As demonstrated in the following sections, a precisely balanced pulse energy configuration across colors effectively suppresses this splitting phenomenon, thereby improving temporal structure and synchronization.

\begin{figure*}[bt]
\includegraphics[width=0.5\textwidth]{fig4.pdf}
\caption{\label{fig:figure_op}(a) The spectra of the multi-color XFEL pulses at the undulator module exit. (b) The various FEL power profiles at the undulator module exit with a time delay of 1 ps. (c) Undulator parameter setting in the whole beamline.}
\end{figure*}

Figure~\ref{fig:figure_op_com}(a) (solid line) shows that the bunching factor in the first sub-undulator $U_{A,1}$ remains nearly zero, indicating the absence of significant microbunching structure in the electron beam. When optimal dispersive strengths $R_{56}^1= 1.84  \ \mu m$, $R_{56}^2= 1.85  \ \mu m$, $R_{56}^3= 1.74  \ \mu m$, and $R_{56}^4= 1.15\ \mu m$ are implemented during the progression through the sub-undulator $U_{A,5}$, the bunching factor undergoes four sharp vertical transitions. These transitions result in a dramatic increase in the bunching factor from the shot-noise level to approximately 0.54. These distinct jumps, under the four-stage OK configuration, signify a substantial enhancement in microbunching. The phenomenon highlights the effectiveness of using four optimized parameters $R_{56}$ to enhance energy modulation and convert it into density modulation four times within the undulator module $A_{5}$. Consequently, this process significantly enhances the FEL gain of the radiation at $\lambda_1$ = 1.8 nm. Between sub-undulators $U_{A,5}$ and $U_{B,1}$, the bunching factor gradually decays and stabilizes at lower values due to microbunching disruption by the delay chicane 1 (delay = 1 ps). Subsequent modules operating in the five-stage OK mode ($B_6$, $C_6$, $D_6$) display identical behavior: each vertical transition represents a phase of microbunching enhancement. Notably, the microbunching is disrupted by delay chicanes 2 and 3, as observed between $U_{B,6}$ and $U_{C,1}$, as well as between $U_{C,6}$ and $U_{D,1}$.

\begin{figure}[bt]
	\includegraphics[width=0.42\textwidth]{fig5.pdf}
	\caption{\label{fig:figure_op_com}Evolution of the slice-averaged bunching factor (a), slice-averaged energy spread (b), and FEL pulse energy (c) along the beamline (z-direction), comparing two configurations: (1) Without OK (dot-dashed plot, plain SASE); (2) With OK (solid plot) and based on three-color X-ray pulses that include OK, with the fourth pulse employing SASE mode (gray dashed plot).}
\end{figure}
As the beam energy spread increases along the undulator modules in Fig.~\ref{fig:figure_op_com}(b), the condition $\sigma_\delta\ll\rho$ is progressively violated, leading to declining pulse energies for the different X-ray wavelengths: 30.4 $\mu J$ (1.8 nm), 25.6 $\mu J$ (2.1 nm), 1.9 $\mu J$ (2.4 nm), and 2.9 $\mu J$ (2.7 nm). This behavior is consistent with the idealized sensitivity analysis shown in Fig.~\ref{fig:re-beam-espread}.
In contrast, when OK chicanes are removed, the corresponding pulse energies drop to 0.2, 4.9, 1.7, and 5.5 $\mu J$. 
The multi-stage OK enhancement is most effective for the first three pulses.  For the fourth pulse, the case with OK does not yield significant FEL improvement due to its larger initial energy spread, compared to the scenario without OK (dot-dashed line), which exhibits a more gradual energy spread growth and thus a relatively smaller initial energy spread. However, when the first three pulses are enhanced by OK while the fourth pulse reverts to normal SASE (gray dashed line), the output performance deteriorates significantly. This indicates that the OK mechanism still plays a crucial role in accelerating the FEL gain process across different wavelengths, thereby enabling four-color XFEL pulse generation within shorter undulator modules. The corresponding output parameters, including peak power, pulse length, and pulse energy, are summarized in Table~\ref{table-ok/sase}.

\begin{table*}
\caption{\label{table-ok/sase} Main output parameters of the four XFEL pulses. Values in each entry are given as \textbf{OK-SASE} / SASE.}
\begin{ruledtabular}
\begin{tabular}{lcccc}
 & First pulse & Second pulse & Third pulse & Fourth pulse \\
\hline
Peak power (GW) 
& \textbf{1.72}/0.02 
& \textbf{2.55}/0.42 
& \textbf{0.53}/0.14 
& \textbf{0.42}/0.75 \\
Pulse length (fs) 
& \textbf{31.6}/44.7 
& \textbf{66.5}/39.0 
& \textbf{71.7}/53.1 
& \textbf{70.2}/48.1 \\

Pulse energy ($\mu$J) 
& \textbf{30.4}/0.2 
& \textbf{25.6}/9.2 
& \textbf{1.9}/1.7 
& \textbf{2.9}/5.5 \\
\end{tabular}
\end{ruledtabular}
\end{table*}
For each wavelength of the X-ray pulse, the $R_{56}$ parameters were sequentially scanned to determine their optimal values by using the differential evolution algorithm. Figure~\ref{fig:figure_op_all} shows the optimal $R_{56}$ configuration of dispersive chicanes in the whole beamline. The optimal $R_{56}$ decreases progressively for each wavelength. This indicates that the strength of the dispersion weakens, while the energy modulation amplitude becomes increasingly larger than the intrinsic energy spread. This trend is dominated by the Bessel function, as described in Fig.~\ref{fig:figure_dif_ener}. Minimal variation is observed in the values of $R_{56}$ for the last two radiation wavelengths. This suggests a diminishing influence of microbunching effects on the electron beam as the energy spread increases. These results demonstrate that the differential evolution algorithm successfully maximized the FEL gain, identifying the optimal values $R_{56}$  for each configuration.
\begin{figure}[bt]
	\includegraphics[width=0.42\textwidth]{fig6.pdf}
	\caption{\label{fig:figure_op_all}Parameter scans of the all-stage dispersive chicanes' $R_{56}$ were performed for different radiation wavelengths. The parameters for the preceding stages were determined sequentially based on prior optimization results. (a) For a radiation wavelength of 1.8 nm, a four-stage OK configuration was used. (b)-(d) For radiation wavelengths of 2.1, 2.4, and 2.7 nm, five-stage OK configurations were employed.}
\end{figure}
\section{\label{sec4}Multi-frame diffraction imaging}
It is necessary to characterize the statistical properties and main parameters of the generated four-color XFEL pulses under realistic operating conditions. In particular, the pulse-energy balance among different colors, the shot-to-shot stability inherent to the SASE process, and the transverse radiation properties play a crucial role in determining the applicability of the source for downstream experiments.

Since the $R_{56}$ values were fully optimized in the previous simulation, the first two colors exhibited substantially higher pulse energies, at the expense of the last two colors, which became relatively weak. To obtain a more evenly tailored pulse-energy distribution among the four colors, additional fine-tuning of the $R_{56}$ values of the dispersive chicanes in the undulator modules was performed. To account for shot-to-shot fluctuations inherent to SASE, the pulse energies were evaluated based on 50-shot simulations. The results, shown in Fig.~\ref{fig:figure_ba}(a), demonstrate the overall stability of the FEL performance, whereas Figs.~\ref{fig:figure_ba}(b) and (c) present a representative single shot illustrating the typical pulse characteristics. Figure~\ref{fig:figure_tra} displays the corresponding transverse power density distributions of the four-color X-ray pulses, from which the rms transverse spot sizes are extracted. The main FEL properties of this representative shot are summarized in Table~\ref{table2}, including pulse energy, pulse length (FWHM), spectral bandwidth (FWHM), transverse spot size (rms), and transverse divergence (rms). These parameters serve as the input conditions for the subsequent X-ray transport simulations. 

\begin{figure}[bt]
	\includegraphics[width=0.4\textwidth]{fig7.pdf}
	\caption{\label{fig:figure_ba}(a) Pulse energies of the four XFEL pulses from 50-shot simulations. (b) Spectra and (c) power profiles of the four-color XFEL pulses at the exit of the undulator modules for a representative single shot.
 }
\end{figure}

\begin{table*}
\caption{\label{table2}FEL performance parameters for pulses at different wavelengths.}
\begin{ruledtabular}
\begin{tabular}{lcccc}
& First pulse & Second pulse  & Third pulse & Fourth pulse   \\
\hline
Pulse energy ($\mu J$)   &11.3  & 8.3  &9.0  &9.2   \\
Pulse length (fs)   &34.9  &46.3  &66.6  &63.3   \\
Bandwidth (nm)               &0.009  & 0.015  &0.024  &0.026   \\
$\sigma_x$ ($\mu m$)         &25.48   & 34.65   &31.59   & 44.84  \\
$\sigma_y$  ($\mu m$)        &22.42   & 24.46   &26.50   &29.55  \\
$\sigma'_x$ ($\mu rad$)      &7.66    & 8.06    &8.72    & 9.59  \\
$\sigma'_y$ ($\mu rad$)      &6.45    & 7.65    &7.64   & 8.84  \\
\end{tabular}
\end{ruledtabular}
\end{table*}

\begin{figure}[bt]
	\includegraphics[width=0.42\textwidth]{fig8.pdf}
	\caption{\label{fig:figure_tra}Transverse intensity profiles of the four-color XFEL pulses at the exit of the undulator module for a representative shot: (a) first pulse, (b) second pulse, (c) third pulse, and (d) fourth pulse.}
\end{figure}


To assess the applicability of this radiation for scattering experiments, its spectroscopic performance is further investigated by analyzing the spectral distribution. In this analysis, neither the dynamical response of the sample nor the temporal structure of the multi-color XFEL pulses is considered, allowing the focus to be placed solely on the intrinsic spectral properties of the radiation. An X-ray grating with 2400~lines/mm is employed, with the beam incident on the grating surface at a grazing angle of $5^\circ$. Only the first diffraction order is taken into account, and a detector is placed at the image plane corresponding to the central wavelength of the FEL radiation to characterize the resulting spectral distribution.


The grating-object distance is set to 5 m, and the image distance is 2 m. Given that our radiation comprises four distinct wavelengths — 1.8, 2.1, 2.4, and 2.7 nm — the distances traveled by each wavelength to reach the grating are 97.07, 66.38, 35.69, and 5 m. Therefore, the footprint of radiation incident on the grating surface corresponds to each wavelength is 1.6 mm for $\lambda_1$, 1.2 mm for $\lambda_2$, 0.7 mm for $\lambda_3$, 0.1 mm for $\lambda_4$. The corresponding grating resolution $R$ values were determined to be 3840, 2880, 1680, and 240, with minimum wavelength resolution $\Delta \lambda_{min}$ equal to 0.47, 0.73, 1.4, and 11.3 pm, respectively. The theoretical resolution limit of the grating for each wavelength is significantly smaller than the FWHM bandwidth of each wavelength. Under ideal conditions, the grating should theoretically be capable of resolving four-color XFEL pulses with wavelength differences of 0.3 nm.

\begin{figure}[bt]
	\includegraphics[width=0.4\textwidth]{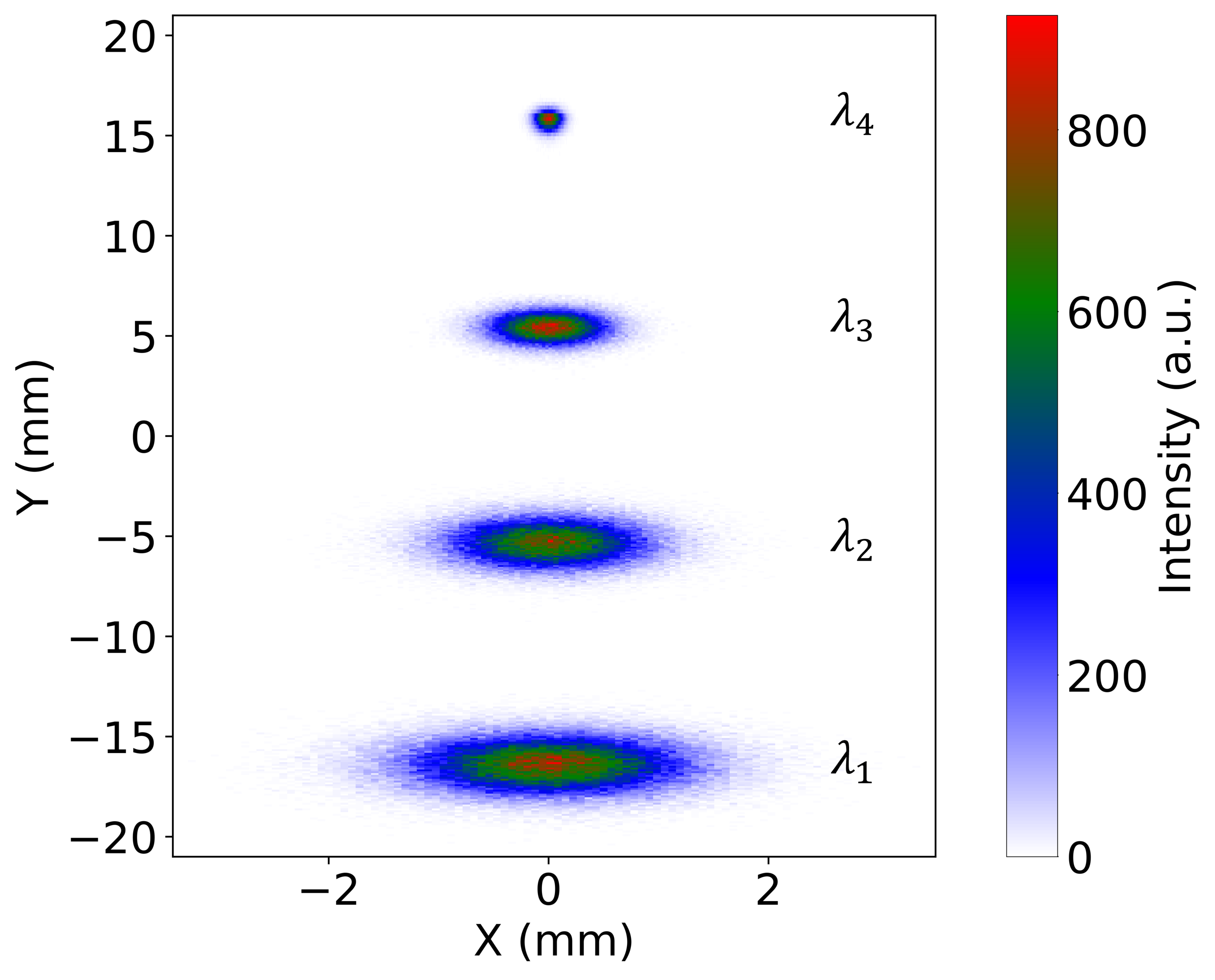}
	\caption{\label{fig:figure_diff}Distinguishable diffraction images were clearly resolved in the simulated image plane, centered around $\lambda_{0}$, with wavelengths of 1.8, 2.1, 2.4, and 2.7 nm arranged from bottom to top.}
\end{figure}

The Shadow 4 \cite{2025shadow} optical tracing code framework was employed for optical tracing simulations of the aforementioned grating system. This framework is commonly used in the field of synchrotron radiation optics.
Consequently, the relative diffraction angle differences from central wavelength ($\lambda_{0}$ = 2.25 nm) are $0.471^\circ$, $0.154^\circ$, $0.151^\circ$, $0.444^\circ$. The distances of these wavelengths from $\lambda_{0}$ are 16.45, 5.37, 5.27, and 15.52 mm, respectively, as shown in Fig.~\ref{fig:figure_diff}. These results could be recorded using a large-area X-ray CCD detector\cite{2021GannNIST} ($4096 \times4096$ pixels, pixel size  $15\times 15\ \mu m ^2$). Thus, this simulation process of X-ray transport design validated the feasibility of the multi-color X-ray pulse diffraction imaging scheme. More importantly, realizing such single-shot, multi-frame probing within the SHINE relies on the introduction of a multi-stage OK configuration, which substantially reduces the effective undulator length required for each color. If repeated pump–probe delay scans are needed, the high repetition rate of SHINE can support this requirement by providing sufficient photon statistics.

\section{\label{sec5}Conclusion and discussion}

A novel single-shot multi-frame X-ray diffraction imaging scheme is proposed by theoretical analysis and start-to-end simulation. The scheme employs the whole electron bunch in a multi-stage OK-SASE configuration with split undulators to generate four-color XFEL pulses featuring independently tunable wavelength separations and controllable inter-pulse delays, while reducing the required undulator length. Delay chicanes placed in the gaps between undulator modules operating at different resonant wavelengths enable time delays on the order of one hundred femtoseconds, extendable into the picosecond regime. The feasibility of the imaging process is confirmed by X-ray transport simulations, which demonstrate that the resulting diffraction signals from distinct probe pulses can be spatially resolved and recorded simultaneously, thereby validating the scheme’s capability for capturing ultrafast dynamics in a single shot.


Furthermore, the capability to flexibly generate wavelengths of interest, such as 1.8, 2.1, 2.4, and 2.7~nm, with a tunable separation (e.g., 0.3~nm) plays an important role in optimizing both the contrast and spatial resolution of diffraction imaging. This flexibility is particularly advantageous for probing elemental absorption edges, for example, the oxygen K-edge at 529~eV. Moreover, extending the scheme to other soft X-ray regions, such as around 6~nm covering the carbon K-edge at 284~eV, further broadens its applicability to a wide range of scientific problems. 
In addition, the achievable temporal resolution is on the order of 100 fs, mainly limited by the XFEL pulse length, while this pulse length can be adjusted through accelerator operating modes. Under more advanced short-pulse operating conditions, it could be reduced to several tens of femtoseconds or even approach 10 fs level, thereby further improving the temporal resolution.
The pulse energy of each color can be effectively controlled by precisely adjusting the $R_{56}$ values of the dispersive chicanes. Balanced pulse energy is about 10~$\mu J$ per color, which remains very low compared with standard single-color operation at existing XFEL facilities. Despite this limitation, such tunability may still be useful for distributing the available photon flux among different colors according to experimental needs and for reducing radiation damage to fragile samples.
Since all pulses are generated from a single electron bunch, independent control of their temporal profiles is inherently limited, and residual energy modulation may induce pulse splitting in subsequent undulator modules. This effect may be mitigated by reducing the radiation intensity in upstream stages while gradually increasing it in downstream stages, or by enlarging the wavelength separation between adjacent pulses. In addition, the residual beam energy chirp of about 0.22\% remains smaller than the FEL gain bandwidth of 0.34\%, so no significant spectral broadening is observed. The implementation of a dechirper would allow more precise manipulation of the longitudinal phase space of the electron beam, offering further flexibility in tailoring pulse characteristics.

The multi-stage OK configuration can also be extended to other multi-color approaches, such as fresh-slice operation, where it can further enhance FEL gain and enable higher peak power or shorter pulses.  In particular, under low-charge operation, where fresh-slice manipulation becomes challenging, the full-beam scheme combined with OK may still provide a practical route to two-color FEL generation. Overall, the proposed scheme is expected to provide a better foundation for designing sample-specific light sources tailored to dynamic imaging experiments for XFEL users, with its multi-color X-ray pulses enabling applications such as multi-color pump-probe experiments, nonlinear X-ray spectroscopy, multi-wavelength anomalous diffraction, and other multi-wavelength techniques. Collectively, these capabilities provide a robust framework for exploring microscopic mechanisms in ultrafast science.
\begin{acknowledgments}
We extend our gratitude to Eduard Prat, Sven Reiche, Manuel Sanchez del Rio, Zipeng Liu, and Zheng Qi for their insightful discussions and keen interest in this work. J. Q. Y. would like to thank the support from the junior visiting professor program of Peng Huanwu Innovation Research Center for Theoretical Physics in 2025. This work was supported by the Natural Science Foundation of China (Grant Nos. 12175058, 12125508, 12541503, 12575251), National Science Fund of Hunan Province for Distinguished Young Scholars No. 2024JJ2009, the National Key Research and Development Program of China (2024YFA1612104), Beijing Distinguished Young Scientist Program and National Grand Instrument Project (Grant No. SQ2019YFF01014400), Shanghai Pilot Program for Basic Research – Chinese Academy of Sciences, Shanghai Branch (JCYJ-SHFY-2021-010), and China Postdoctoral Science Foundation (2025M770914).
\end{acknowledgments}

\nocite{*}
\raggedbottom

\bibliography{manuscript}

@article{2007huangReviewXrayFreeelectron,
  title = {Review of X-Ray Free-Electron Laser Theory},
  author = {Huang, Zhirong and Kim, Kwang-Je},
  year = {2007},
  month = mar,
  journal = {Physical Review Special Topics - Accelerators and Beams},
  volume = {10},
  number = {3},
  pages = {034801} 
}

@article{2016pellegriniPhysicsXrayFreeelectron,
  title = {The Physics of X-Ray Free-Electron Lasers},
  author = {Pellegrini, C. and Marinelli, A. and Reiche, S.},
  year = {2016},
  month = mar,
  journal = {Reviews of Modern Physics},
  volume = {88},
  number = {1},
  pages = {015006},
}

@article{2021Features,
  title={Features and futures of X-ray free-electron lasers.},
  author={ Huang, Nanshun  and  Deng, Haixiao  and  Liu, Bo  and  Wang, Dong  and  Zhao, Zhentang },
  journal={Innovation (Cambridge (Mass.))},
  number={2},
  year={2021},
  volume ={2},
  pages = {100097}
}

@article{2017chemattarFemtosecondXraySpectroscopya,
  title = {Femtosecond X-Ray Spectroscopy of an Electrocyclic Ring-Opening Reaction},
  author = {Attar, Andrew R. and Bhattacherjee, Aditi and Pemmaraju, C. D. and Schnorr, Kirsten and Closser, Kristina D. and Prendergast, David and Leone, Stephen R.},
  year = {2017},
  month = apr,
  journal = {Science},
  volume = {356},
  number = {6333},
  pages = {54--59}
}

@article{2018chemincigrucciImpulsiveUVpumpXraya,
  title = {Impulsive {{UV-pump}}/{{X-ray}} Probe Study of Vibrational Dynamics in Glycine},
  author = {Mincigrucci, Riccardo  and  Kowalewski, Markus  and Rouxel, Jérémy R. and  Bencivenga, Filippo  and  Mukamel, Shaul  and  Masciovecchio, Claudio},
  year = {2018},
  month = oct,
  journal = {Scientific Reports},
  volume = {8},
  number = {1},
  pages = {15466}
}

@article{2019berrahFemtosecondresolvedObservationFragmentation,
  title = {Femtosecond-Resolved Observation of the Fragmentation of Buckminsterfullerene Following {{X-ray}} Multiphoton Ionization},
  author = {Berrah, N.  and  Sanchez-Gonzalez, A.  and  Jurek, Z.  and  Obaid, R.  and  Santra, R.},
  year = {2019},
  month = dec,
  journal = {Nature Physics},
  volume = {15},
  number = {12},
  pages = {1279--1283}
}

@article{2016ferrariWidelyTunableTwocolour,
  title = {Widely Tunable Two-Colour Seeded Free-Electron Laser Source for Resonant-Pump Resonant-Probe Magnetic Scattering},
  author = {Ferrari, Eugenio  and  Spezzani, Carlo  and  Fortuna, Franck  and  Delaunay, Renaud  and  Sacchi, Maurizio},
  year = {2016},
  month = jan,
  journal = {Nature Communications},
  volume = {7},
  number = {1},
  pages = {10343},
}

@article{2013allariaTwocolourPumpProbe,
  title = {Two-Colour Pump-Probe Experiments with a Twin-Pulse-Seed Extreme Ultraviolet Free-Electron Laser},
  author = {Allaria, E.  and  Bencivenga, F.  and  Borghes, R.  and  Capotondi, F.  and  Zangrando, M.},
  year = {2013},
  month = sep,
  journal = {Nature Communications},
  volume = {4},
  number = {1},
  pages = {2476},
  issn = {2041-1723},
 
}

@article{2016Observation,
  title={Observation of femtosecond X-ray interactions with matter using an X-ray-X-ray pump-probe scheme.},
  author={ Inoue, Ichiro  and  Inubushi, Yuichi  and  Sato, Takahiro  and  Tono, Kensuke  and  Katayama, Tetsuo  and  Kameshima, Takashi  and  Ogawa, Kanade  and  Togashi, Tadashi  and  Owada, Shigeki  and  Amemiya, Yoshiyuki },
  journal={Proceedings of the National Academy of Sciences},
  pages={201516426},
  year={2016},
}

@article{2012materipfauUltrafastOpticalDemagnetization,
  title = {Ultrafast Optical Demagnetization Manipulates Nanoscale Spin Structure in Domain Walls},
  author = {Pfau, B. and others},
  year = {2012},
  month = oct,
  journal = {Nature Communications},
  volume = {3},
  number = {1},
  pages = {1100}
}

@article{2017materiferrariElementSelectiveProbea,
  title = {Element {{Selective Probe}} of the {{Ultra-Fast Magnetic Response}} to an {{Element Selective Excitation}} in {{Fe-Ni Compounds Using}} a {{Two-Color FEL Source}}},
  author = {Ferrari, Eugenio and others},
  year = {2017},
  month = jan,
  journal = {Photonics},
  volume = {4},
  number = {1},
  pages = {6},
}

@article{2018densepontiusProbingNonequilibriumTransienta,
  title = {Probing the Non-Equilibrium Transient State in Magnetite by a Jitter-Free Two-Color {{X-ray}} Pump and {{X-ray}} Probe Experiment},
  author = {Pontius, N.  and  Beye, M.  and  Trabant, C.  and  Mitzner, R.  and  Sorgenfrei, F.  and  Kachel, T.  and Wöstmann, M. and  Roling, S.  and  Zacharias, H.  and  Ivanov, R.},
  year = {2018},
  month = sep,
  journal = {Structural Dynamics},
  volume = {5},
  number = {5},
  pages = {054501},
  issn = {2329-7778},
 
}

@article{2022densejohnsonUltrafastXrayImaginga,
  title = {Ultrafast {{X-ray}} Imaging of the Light-Induced Phase Transition in {{VO2}}},
  author = {Johnson, Allan S. and others},
  year = {2022},
  month = dec,
  journal = {Nature Physics},
}

@article{2016plasmagorkhoverFemtosecondNanometreVisualization,
  title = {Femtosecond and Nanometre Visualization of Structural Dynamics in Superheated Nanoparticles},
  author = {Gorkhover, Tais and others},
  year = {2016},
  month = feb,
  journal = {Nature Photonics},
  volume = {10},
  number = {2},
  pages = {93--97},
}

@article{2023plasmahoeingTimeresolvedSingleparticleXraya,
  title = {Time-Resolved Single-Particle x-Ray Scattering Reveals Electron-Density as Coherent Plasmonic-Nanoparticle-Oscillation Source},
  author = {Hoeing, D. and others},
  year = {2023},
  month = jul,
  journal = {Nano Letters},
  volume = {23},
  number = {13},
 
  
  pages = {5943--5950},
}

@article{2021chenobservation,
  title={Observation of a highly conductive warm dense state of water with ultrafast pump--probe free-electron-laser measurements},
  author={Chen, Z and others},
  journal={Matter and Radiation at Extremes},
  volume={6},
  number={5},
  year={2021},
  publisher={AIP Publishing}
}

@article{2011biocalemanSimulationsRadiationDamagea,
  title = {Simulations of Radiation Damage in Biomolecular Nanocrystals Induced by Femtosecond {{X-ray}} Pulses},
  author = {Caleman, Carl and Bergh, Magnus and Scott, Howard A. and Spence, John C.H. and Chapman, Henry N. and T{\^i}mneanu, Nicu{\c s}or},
  year = {2011},
  month = sep,
  journal = {Journal of Modern Optics},
  volume = {58},
  number = {16},
  pages = {1486--1497},
  issn = {0950-0340, 1362-3044},
}

@article{2017biowolfProbingUltrafastPpa,
  title = {Probing Ultrafast {$\Pi\pi$}*/N{$\pi$}* Internal Conversion in Organic Chromophores via {{K-edge}} Resonant Absorption},
  author = {Wolf, T. J. A. and others},
  year = {2017},
  month = jun,
  journal = {Nature Communications},
  volume = {8},
  number = {1},
  pages = {29},
}

@article{2014nakagawaSequentiallyTimedAlloptical,
  title = {Sequentially Timed All-Optical Mapping Photography ({{STAMP}})},
  author = {Nakagawa, K. and Iwasaki, A. and Oishi, Y. and Horisaki, R. and Tsukamoto, A. and Nakamura, A. and Hirosawa, K. and Liao, H. and Ushida, T. and Goda, K. and Kannari, F. and Sakuma, I.},
  year = {2014},
  month = sep,
  journal = {Nature Photonics},
  volume = {8},
  number = {9},
  pages = {695--700}
}

@article{2023saikiSingleshotOpticalImaging,
  title = {Single-Shot Optical Imaging with Spectrum Circuit Bridging Timescales in High-Speed Photography},
  author = {Saiki, Takao and others},
  year = {2023},
  month = dec,
  journal = {Science Advances},
  volume = {9},
  number = {51},
  pages = {eadj8608},
  
}

@article{2015marinelliHighintensityDoublepulseXray,
  title = {High-Intensity Double-Pulse {{X-ray}} Free-Electron Laser},
  author = {Marinelli, A. and others},
  year = {2015},
  month = mar,
  journal = {Nature Communications},
  volume = {6},
  number = {1},
  pages = {6369},
}

@article{2022deckerTunableXrayFree,
  title = {Tunable X-Ray Free Electron Laser Multi-Pulses with Nanosecond Separation},
  author = {Decker, Franz-Josef and others},
  year = {2022},
  month = feb,
  journal = {Scientific Reports},
  volume = {12},
  number = {1},
  pages = {3253}
}

@article{2021s.bettoniExperimentalDemonstrationTwocolor,
  title = {Experimental Demonstration of Two-Color x-Ray Free-Electron-Laser Pulses via Wakefield Excitation},
  author = {{S. Bettoni} and Craievich, P. and Dax, A. and Ganter, R. and Guetg, M. W. and Huppert, M. and Marcellini, F. and Neto Pestana, R. and Reiche, S. and Prat, E. and Trisorio, A. and Vicario, C. and Lutman, A. A.},
  year = {2021},
  month = aug,
  journal = {Physical Review Accelerators and Beams},
  volume = {24},
  number = {8},
  pages = {082801}
}

@article{2019saahernandezGenerationTwocolorXray,
  title = {Generation of Two-Color x-Ray Free-Electron-Laser Pulses from a Beam with a Large Energy Chirp and a Slotted Foil},
  author = {Sa{\'a} Hern{\'a}ndez, {\'A}ngela and Prat, Eduard and Reiche, Sven},
  year = {2019},
  month = mar,
  journal = {Physical Review Accelerators and Beams},
  volume = {22},
  number = {3},
  pages = {030702}
}

@article{2020dijkstalDemonstrationTwocolorXray,
  title = {Demonstration of Two-Color x-Ray Free-Electron Laser Pulses with a Sextupole Magnet},
  author = {Dijkstal, P. and Malyzhenkov, A. and Reiche, S. and Prat, E.},
  year = {2020},
  month = mar,
  journal = {Physical Review Accelerators and Beams},
  volume = {23},
  number = {3},
  pages = {030703}
}

@article{2020malyzhenkovSingleTwocolorAttosecond,
  title = {Single- and Two-Color Attosecond Hard x-Ray Free-Electron Laser Pulses with Nonlinear Compression},
  author = {Malyzhenkov, Alexander and Arbelo, Yunieski P. and Craievich, Paolo and Dijkstal, Philipp and Ferrari, Eugenio and Reiche, Sven and Schietinger, Thomas and Jurani{\'c}, Pavle and Prat, Eduard},
  year = {2020},
  month = oct,
  journal = {Physical Review Research},
  volume = {2},
  number = {4},
  pages = {042018}
}

@article{2021vicarioTwocolorXrayFreeelectron,
  title = {Two-Color x-Ray Free-Electron Laser by Photocathode Laser Emittance Spoiler},
  author = {Vicario, Carlo and Bettoni, Simona and Lutman, Alberto and Dax, Andreas and Huppert, Martin and Trisorio, Alexandre},
  year = {2021},
  month = jun,
  journal = {Physical Review Accelerators and Beams},
  volume = {24},
  number = {6},
  pages = {060703}
}

@article{2013lutmanExperimentalDemonstrationFemtosecond,
  title = {Experimental {{Demonstration}} of {{Femtosecond Two-Color X-Ray Free-Electron Lasers}}},
  author = {Lutman, A. A. and Coffee, R. and Ding, Y. and Huang, Z. and Krzywinski, J. and Maxwell, T. and Messerschmidt, M. and Nuhn, H.-D.},
  year = {2013},
  month = mar,
  journal = {Physical Review Letters},
  volume = {110},
  number = {13},
  pages = {134801}
}

@article{2013haraTwocolourHardXray,
  title = {Two-Colour Hard {{X-ray}} Free-Electron Laser with Wide Tunability},
  author = {Hara, Toru and others},
  year = {2013},
  month = dec,
  journal = {Nature Communications},
  volume = {4},
  number = {1},
  pages = {2919}
}

@article{2023choGenerationTimesynchronizedTwocolor,
  title = {Generation of Time-Synchronized Two-Color {{X-ray}} Free-Electron Laser Pulses Using Phase Shifters},
  author = {Cho, Myung-Hoon and Kang, Teyoun and Yang, Haeryong and Kim, Gyujin and Kwon, Seong-Hoon and Moon, Kook-Jin and Nam, Inhyuk and Min, Chang-Ki and Heo, Hoon and Kim, Changbum and Kang, Heung-Sik and Shim, Chi Hyun},
  year = {2023},
  month = aug,
  journal = {Scientific Reports},
  volume = {13},
  number = {1},
  pages = {13786}
}

@article{2016lutmanFreshsliceMulticolourXray,
  title = {Fresh-Slice Multicolour {{X-ray}} Free-Electron Lasers},
  author = {Lutman, Alberto A. and Maxwell, Timothy J. and MacArthur, James P. and Guetg, Marc W. and Berrah, Nora and Coffee, Ryan N. and Ding, Yuantao and Huang, Zhirong and Marinelli, Agostino and Moeller, Stefan and Zemella, Johann C. U.},
  year = {2016},
  month = nov,
  journal = {Nature Photonics},
  volume = {10},
  number = {11},
  pages = {745--750}
}

@article{2016reicheTwocolorOperationFreeelectron,
  title = {Two-Color Operation of a Free-Electron Laser with a Tilted Beam},
  author = {Reiche, Sven and Prat, Eduard},
  year = {2016},
  month = jul,
  journal = {Journal of Synchrotron Radiation},
  volume = {23},
  number = {4},
  pages = {869--873}
}

@article{2018guetgDispersionBasedFreshSliceScheme,
  title = {Dispersion-{{Based Fresh-Slice Scheme}} for {{Free-Electron Lasers}}},
  author = {Guetg, Marc W. and Lutman, Alberto A. and Ding, Yuantao and Maxwell, Timothy J. and Huang, Zhirong},
  year = {2018},
  month = jun,
  journal = {Physical Review Letters},
  volume = {120},
  number = {26},
  pages = {264802}
}

@article{2018chaoControlLasingSlice,
  title = {Control of the {{Lasing Slice}} by {{Transverse Mismatch}} in an {{X-Ray Free-Electron Laser}}},
  author = {Chao, Yu-Chiu and Qin, Weilun and Ding, Yuantao and Lutman, Alberto A. and Maxwell, Timothy},
  year = {2018},
  month = aug,
  journal = {Physical Review Letters},
  volume = {121},
  number = {6},
  pages = {064802}
}

@article{2022pratWidelyTunableTwocolor,
  title = {Widely Tunable Two-Color x-Ray Free-Electron Laser Pulses},
  author = {Prat, Eduard and others},
  year = {2022},
  month = may,
  journal = {Physical Review Research},
  volume = {4},
  number = {2},
  pages = {L022025}
}

@article{2024Millijoule,
  title={Millijoule Femtosecond X-Ray Pulses from an Efficient Fresh-Slice Multistage Free-Electron Laser},
  author={ Wang, Guanglei  and  Dijkstal, Philipp  and  Reiche, Sven  and  Schnorr, Kirsten  and  Prat, Eduard },
  journal={Physics Review Letter},
  volume={132},
  number={3},
  pages={6},
  year={2024},
}

@article{2011guntherSequentialFemtosecondXray,
  title = {Sequential Femtosecond {{X-ray}} Imaging},
  author = {G{\"u}nther, C. M. and Pfau, B. and Mitzner, R. and Siemer, B. and Roling, S. and Zacharias, H. and Kutz, O. and Rudolph, I. and Schondelmaier, D. and Treusch, R. and Eisebitt, S.},
  year = {2011},
  month = feb,
  journal = {Nature Photonics},
  volume = {5},
  number = {2},
  pages = {99--102},
}

@article{1992bonifacioTheoryHighgainOptical,
  title = {Theory of the High-Gain Optical Klystron},
  author = {Bonifacio, R. and Corsini, R. and Pierini, P.},
  year = {1992},
  month = mar,
  journal = {Physical Review A},
  volume = {45},
  number = {6},
  pages = {4091--4096}
}

@article{2001Dispersively,
  title={Dispersively enhanced bunching in high-gain free-electron lasers},
  author={ Neil, G. R  and  Freund, H. P },
  journal={Nuclear Inst \& Methods in Physics Research A},
  volume={475},
  number={1-3},
  pages={381-384},
  year={2001},
}

@misc{2003saldinFreeElectronLaser,
  title = {The {{Free Electron Laser Klystron Amplifier Concept}}},
  author = {Saldin, E. L. and Schneidmiller, E. A. and Yurkov, M. V.},
  year = {2003},
  month = aug,
  number = {arXiv:physics/0308060},
  
  publisher = {arXiv}
}

@article{2006Optical,
  title={Optical klystron enhancement to self-amplified spontaneous emission free electron lasers},
  author={ Ding, Yuantao  and  Emma, Paul  and  Huang, Zhirong  and  Kumar, Vinit },
  journal={Physical Review Accelerators and Beams},
  volume={9},
  number={7},
  pages={70702-70702},
  year={2006},
}

@article{2015pencoExperimentalDemonstrationEnhanced,
  title = {Experimental {{Demonstration}} of {{Enhanced Self-Amplified Spontaneous Emission}} by an {{Optical Klystron}}},
  author = {Penco, G. and Allaria, E. and De Ninno, G. and Ferrari, E. and Giannessi, L.},
  year = {2015},
  month = jan,
  journal = {Physical Review Letters},
  volume = {114},
  number = {1},
  pages = {013901}
}

@article{2017pencoOpticalKlystronEnhancement,
  title = {Optical {{Klystron Enhancement}} to {{Self Amplified Spontaneous Emission}} at {{FERMI}}},
  author = {Penco, Giuseppe and Allaria, Enrico and De Ninno, Giovanni and Ferrari, Eugenio and Giannessi, Luca and Roussel, El{\'e}onore and Spampinati, Simone},
  year = {2017},
  month = mar,
  journal = {Photonics},
  volume = {4},
  number = {4},
  pages = {15}
}

@article{2021Development,
  title = {Demonstration of a Compact X-Ray Free-Electron Laser Using the Optical Klystron Effect},
  author = {Prat, Eduard and Ferrari, Eugenio and Calvi, Marco and Ganter, Romain and Reiche, Sven and Schmidt, Thomas},
  year = {2021},
  month = oct,
  journal = {Applied Physics Letters},
  volume = {119},
  number = {15},
  pages = {151102},
  issn = {0003-6951},
  }

@inproceedings{1984CollectiveInstabilitiesHighgainbonifacio,
  title = {Collective Instabilities and High-Gain Regime Free Electron Laser},
  booktitle = {{{AIP Conference Proceedings}}},
  author = {Bonifacio, R. and Pellegrini, C. and Narducci, L. M.},
  year = {1984},
  volume = {118},
  pages = {236--259},
  publisher = {AIP},
  
}

@article{1991yuGenerationIntenseUv,
  title = {Generation of Intense Uv Radiation by Subharmonically Seeded Single-Pass Free-Electron Lasers},
  author = {Yu, L. H.},
  year = {1991},
  month = oct,
  journal = {Physical Review A},
  volume = {44},
  number = {8},
  pages = {5178--5193}
}

@article{2021Soft,
  title={Soft X-ray spectrometers based on aperiodic reflection gratings and their application},
  author={ Ragozin, E N  and  Vishnyakov, E A  and  Kolesnikov, A O  and  Pirozhkov, A S  and  Shatokhin, A N },
  journal={Physics-Uspekhi},
  volume={64},
  number={5},
  pages={495-514},
  year={2021},
}

@article{2025BaejDetPy,
  title={DetPy (Differential Evolution Tools): A Python toolbox for solving optimization problems using differential evolution},
  author={Baej Zieliński and  Ciegienny, Szymon  and  Orlicki, Hubert  and  Ksiek, Wojciech },
  journal={SoftwareX},
  volume={29},
  year={2025},
  pages = {102014},
}

@article{2023SHINE,
  title = {Status and Future of the Soft {{X-ray}} Free-Electron Laser Beamline at the {{SHINE}}},
  author = {Liu, Tao and Huang, Nanshun and Yang, Hanxiang and Qi, Zheng and Zhang, Kaiqing and Gao, Zhangfeng and Chen, Si and Feng, Chao and Zhang, Wei and Luo, Hang and Fu, Xiaoxi and Liu, He and Faatz, Bart and Deng, Haixiao and Liu, Bo and Wang, Dong and Zhao, Zhentang},
  year = {2023},
  month = may,
  journal = {Frontiers in Physics},
  volume = {11},
  pages = {1172368},
  issn = {2296-424X}, 
}

@article{2025chenultra,
  title={Ultra-high quality factor and ultra-high accelerating gradient achievements in a 1.3 GHz continuous wave cryomodule},
  author={Chen, Jin-Fang and Zong, Yue and Pu, Xiao-Yun and Xiang, Sheng-Wang and Xing, Shuai and Li, Zheng and Liu, Xu-Ming and Zhai, Yan-Fei and Wu, Xiao-Wei and He, Yong-Zhou and others},
  journal={Nuclear Science and Techniques},
  volume={36},
  number={2},
  pages={25},
  year={2025},
  publisher={Springer}
}

@article{2019yanMultibeamenergyOperationContinuouswave,
  title = {Multi-Beam-Energy Operation for the Continuous-Wave x-Ray Free Electron Laser},
  author = {Yan, Jiawei and Deng, Haixiao},
  year = {2019},
  month = sep,
  journal = {Physical Review Accelerators and Beams},
  volume = {22},
  number = {9},
  pages = {090701}
}

@article{2021yanself,
  title={Self-amplification of coherent energy modulation in seeded free-electron lasers},
  author={Yan, Jiawei and Gao, Zhangfeng and Qi, Zheng and Zhang, Kaiqing and Zhou, Kaishang and Liu, Tao and Chen, Si and Feng, Chao and Li, Chunlei and Feng, Lie and others},
  journal={Physical Review Letters},
  volume={126},
  number={8},
  pages={084801},
  year={2021},
  publisher={APS}
}

@article{2025qifirst,
  title={First Lasing and Stable Operation of a Direct-Amplification Enabled Harmonic Generation Free-Electron Laser},
  author={Qi, Zheng and Liu, Junhao and Ni, Lanpeng and Liu, Tao and Wang, Zhen and Zhang, Kaiqing and Yang, Hanxiang and Gao, Zhangfeng and Huang, Nanshun and Chen, Si and others},
  journal={Physical Review Letters},
  volume={135},
  number={3},
  pages={035001},
  year={2025},
  publisher={APS}
}

@article{2004Astra,
  title={Recent improvements to the ASTRA particle tracking code},
  author={ Flottmann, K.  and  Lidia, S. M.  and  Piot, P. },
  journal={Office of Scientific \& Technical Information Technical Reports},
  volume={5},
  number={3},
  pages={3500-3502},
  year={2004},
}

@article{2001elegant,
  title={Simple method for particle tracking with coherent synchrotron radiation},
  author={ Borland, M },
  journal={Phys.rev.st Accel.beams},
  volume={4},
  number={7},
  pages={289-293},
  year={2001},
}

@article{S1999GENESIS,
  title={GENESIS 1.3: a fully 3D time-dependent FEL simulation code},
  author={S. Reiche},
  journal={Nuclear Instruments \& Methods in Physics Research},
  year={1999}
}

@article{2025shadow,
  title = {{{SHADOW4}}: The Popular Ray Tracing Revived for Evolving Synchrotron Sources in Fourth-Generation Storage Rings},
  shorttitle = {{{SHADOW4}}},
  author = {Del Rio, Manuel Sanchez and {Reyes-Herrera}, Juan and Shi, Xianbo and Rebuffi, Luca},
  year = {2025},
  month = may,
  journal = {Journal of Physics: Conference Series},
  volume = {3010},
  number = {1},
  pages = {012071},
  issn = {1742-6588, 1742-6596},
  
  urldate = {2025-06-17}  
}

@article{2021GannNIST,
  author  = {Eliot Gann and Thomas Crofts and Glenn Holland and Peter Beaucage and Terry McAfee and R. Joseph Kline and Brian A. Collins and Christopher R. McNeill and Daniel A. Fischer and Dean M. DeLongchamp},
  title   = {A {NIST} facility for resonant soft x-ray scattering measuring nano-scale soft matter structure at {NSLS-II}},
  journal = {Journal of Physics: Condensed Matter},
  year    = {2021},
  volume  = {33},
  number  = {16},
 
  issn    = {0953-8984},
  note    = {Published online: 2021-03-10}
}

\end{document}